\documentclass[twocolumn]{revtex4}

\usepackage{graphicx}

\newcommand{\GaMnAs}[1]{\mbox{Ga$_{1-x}$Mn$_x$As}}
\newcommand{\etal}{{\it et al.}}
\newcommand{\CdMnTe}{\mbox{CdMnTe}}

\begin{document}

\title{Optical response of a ferromagnetic/DMS hybrid structure}



\author{P. Redli\'{n}ski}
\email{Pawel.Redlinski.1@nd.edu} \affiliation{Department of Physics,
University of Notre Dame, Notre Dame, IN 46556}

\author{T. Wojtowicz}
\affiliation{Department of Physics, University of Notre Dame, Notre
Dame, IN 46556} \affiliation{Institute of Physics, Polish Academy of
Sciences, Warsaw, Poland}

\author{T. G. Rappoport}
\affiliation{Department of Physics, University of Notre Dame, Notre
Dame, IN 46556}

\author{A. Libal}
\affiliation{Department of Physics, University of Notre Dame, Notre
Dame, IN 46556}

\author{J. K. Furdyna}
\affiliation{Department of Physics, University of Notre Dame, Notre
Dame, IN 46556}

\author{B. Jank\'{o}}
\affiliation{Department of Physics, University of Notre Dame, Notre
Dame, IN 46556}


\date{\today}

\begin{abstract}
We investigate the possibility of using local magnetic fields to
produce \mbox{one-dimensional} traps in hybrid structures for any
quasiparticle possessing spin degree of freedom. We consider a
system composed of a diluted magnetic semiconductor quantum well
buried below a micron-sized ferromagnetic island. Localized magnetic
field is produced by a rectangular ferromagnet in close proximity of
a single domain phase. We make quantitative predictions for the
optical response of the system as a function of distance between the
micromagnet and the quantum well, electronic \mbox{g-factor}, and
thickness of the micromagnet.
\end{abstract}


\maketitle

The use of the spin of quasi-particles, instead of their charge, as
a basis for the operation of a new type of electronic devices has
attracted the attention of a large interdisciplinary research
community. The interest is not only in the phenomenology but also in
the large scale production and applicability of such devices
\cite{Zutic}. In this paper we show that the spin degree of freedom
can be utilized for achieving spatial localization of charged
quasi-particles such as electrons, holes or trions \cite{Wojs1,Ja1},
as well as of neutral complexes such as excitons
\cite{Peeters1,Riva1}. This is of interest for spintronic
applications. In diluted magnetic semiconductors (DMS) like CdMnTe,
due to the exchange interaction between delocalized band electrons
and localized magnetic ions, the presence of a static magnetic field
leads to a giant Zeeman splitting between band states for different
spin components. As a consequence, the effective \mbox{$g$-factor}
for the DMS is very high and temperature dependent
\cite{Gaj1,Furdyna1}.

Here we consider a hybrid structure of a \mbox{CdMnTe/CdMgTe}
quantum well (QW) at tens of nanometers below a rectangular
ferromagnetic island. We find that due to the giant Zeeman
interaction, the non-homogeneous magnetic field produced by the
rectangular ferromagnetic island acts as an effective potential that
can efficiently ``trap'' spin polarized quasi-particles in the QW.
We present quantitative predictions for the optical response of the
DMS where the localization of the quasi-particles is evident. We
also discuss how these predictions are sensitive to the variation of
certain parameters, such as the distance between the QW and the
ferromagnetic island, the thickness of the micromagnet, and the
electronic g-factors (\mbox{$g_e$ - g-factor} of the electron,
\mbox{$g_h$ - g-factor} of the hole).

\begin{figure}[ht]
\includegraphics[width=7.cm]{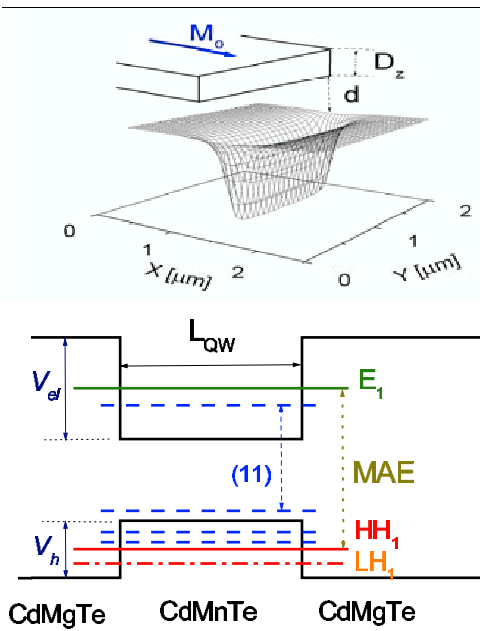}
\caption{Top: z-component of the magnetic field at distance $d$
below magnetic island. The micromagnet is magnetized in the
\mbox{x-direction}. Bottom: with no magnetic island present, we
expect the main absorption edge (MAE) to be between $HH_1$ and $E_1$
states shown by the solid lines. Conduction as well as valence band
edges are plotted for a QW structure without any additional magnetic
field; After the deposition of the island new states appear below
$E_1$ and above $HH_1$ (dashed lines), and each energy state is
\textit{non-degenerate}. For additional explanation see
text.}\label{figQW}
\end{figure}

We consider a typical rectangular micromagnet (iron) of dimensions
$D_x$=6~$\mu$m, $D_y$=2~$\mu$m and $D_z$=0.15~$\mu$m in a single
domain state \cite{Jackson}, with magnetization pointing in the
\mbox{x-direction} \cite{Kossut1,Cywinski,Crowell},
see~Fig.~\ref{figQW}. The single domain state of the micromagnet was
investigated with micro-magnetic simulation using the OOMMF package
made available by NIST \cite{oommf}. In the absence of an external
magnetic field the micromagnet has a multi-domain structure. The
simulation shows that after magnetizing the sample with a field
of~1~T and then reducing the field to~0~T, a value of~0.2~T is
enough to restore a state that is, for our purposes, sufficiently
close to a single domain. Because of the magnetic anisotropy of the
\mbox{$g_h$-factor}, this additional field is unimportant for an
electron in the valence band but can slightly affect electrons in
the conduction band.

Due to the spatial variation of the local magnetic field
(Fig.~\ref{figQW}) \cite{Kossut1, Crowell} and to the presence of
the QW, the quasi-particles are localized below the poles of the
micromagnet \cite{Kossut1}. Localization occurs only in two spatial
dimensions ($x$ and $z$), while the quasi-particles can move
quasi-freely in the third ($y$) direction. As we will show later,
the spin traps and optical response of the quasi-particles are very
sensitive to the value of electron and hole \mbox{g-factors}. In
most of our calculations we use an electron \mbox{g-factor}
\mbox{$g_e$=500} obtained experimentally and reported by Dietl
\etal{} \cite{Dietl1}. Assuming that in CdTe structures the ratio of
exchange interactions $|\beta/\alpha| =4$, we obtain
\mbox{$g_h$=2000}.

\begin{figure}[ht]
\includegraphics[width=7.5cm]{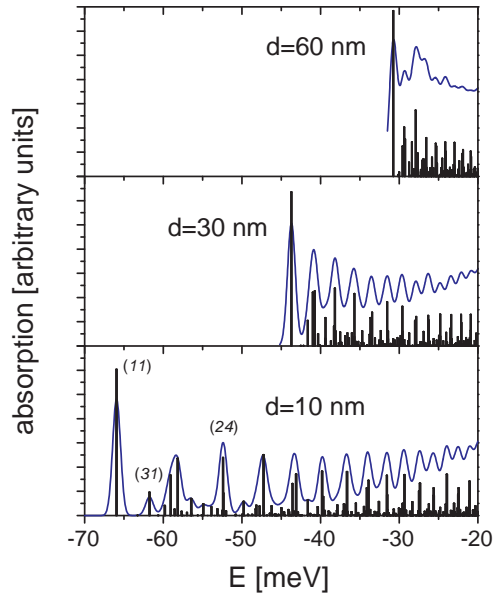}
\caption{Absorption coefficient for three different distances $d$
between the rectangular ferromagnetic island and the QW. The
thickness of the micromagnet is $D_z$=150~nm, and \mbox{$g_e$=500}.
The numbers (nm) correspond to transitions between the $n$th
electron state and the $m$th hole state.}\label{FigRectAbsVsd}
\end{figure}

In Fig.~\ref{figQW} we present a sketch of the energy levels of the
QW for two cases: first, with no ferromagnetic island present; and
second, in the presence of a ferromagnetic island on top of the QW.
With no ferromagnetic island, the main absorption edge (MAE) in the
$\sigma_+$ polarization of light is observed between the states of
the heavy hole $HH_1$ with spin $-3/2$ and the electron $E_1$ with
spin $-1/2$. Due to the absence of magnetic field, $E_1$ and $HH_1$
are two-fold degenerate with respect to the direction of the
pseudo-spins; and the first optical transition in $\sigma_-$
polarization is at the same energy as the $\sigma_+$ transition.
After depositing the micromagnet, we expect new states to appear
below $E_1$ and above $HH_1$. As indicated in Fig.~\ref{figQW},
depending on the parameters, some of the states can be even inside
the QW gap. Note that the edges of the conduction as well as the
valence bands are plotted for the quantum structure without the
effect of the island. The new states are non-degenerate, since the
presence of a local magnetic field lifts the Kramers degeneracy.
Below $E_1$ we have only conduction electron states of spin $-1/2$;
and above $HH_1$, states are built up mainly from valence electrons
with pseudo-spin $-3/2$. Transitions between these new states appear
in the absorption below the main absorption edge according to
selection rules for the $\sigma_{\pm}$ polarizations, as shown in
Fig.~\ref{FigRectAbsVsd} \cite{Madelung}. The label (11) in
Fig.~\ref{FigRectAbsVsd} shows the lowest possible transition
between valence and conduction band states.

\begin{figure}[ht]
\includegraphics[width=7.5cm]{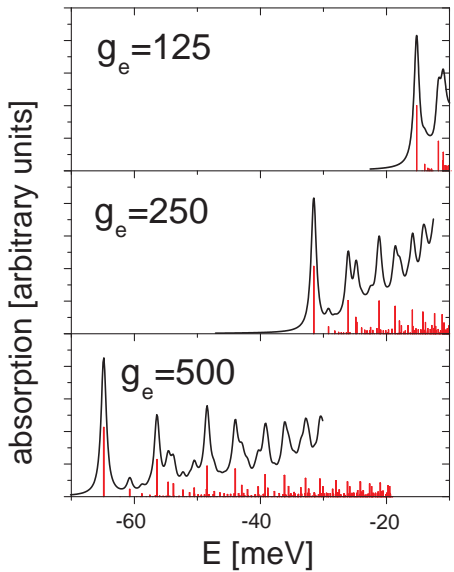}
\caption{Absorption coefficient for different $g_e$-factors for a
rectangular micromagnet. The ratio $|g_h/g_e|$=4 is assumed for the
\CdMnTe{} semiconductor quantum well structure; and we have chosen
$d$=10~nm and $D_z$=150~nm.}\label{FigRectAbsVsG}
\end{figure}

\begin{figure}[ht]
\includegraphics[width=8.0cm]{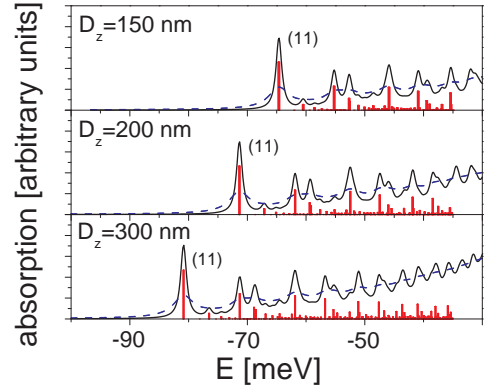}
\caption{Absorption coefficient for different thickness $D_z$ of a
micromagnet. Distance between micromagnet and quantum well $d$=10~nm
and \mbox{$g_e$=500}. Each transition energy is broadened by 1~meV
(solid lines) and 4~meV (dashed lines).}\label{FigRectAbsVsDz}
\end{figure}

For typical quasi-particle masses the  motion of the conduction and
valence electrons is quasi-free in the \mbox{y-direction}. Therefore
we should rather speak  about bands than individual levels. On the
other hand we know that Coulomb interaction will create excitons
which form sharp individual optical lines. Furthermore, it is
assumed that exciton states are formed and the MAE corresponds now
to the 1S exciton transition line in the QW without the island. In
Fig.~\ref{FigRectAbsVsd}, Fig.~\ref{FigRectAbsVsG} and
Fig.~\ref{FigRectAbsVsDz} we take the zero energy at the 1S exciton
main absorption peak.

In our calculations we model valence band electrons using a two-band
Luttinger Hamiltonian \cite{Lutt,Abolfath}, and for the conduction
band electrons we use a quadratic Hamiltonian. Including the Zeeman
interaction, we diagonalize the total Hamiltonian and obtain one
electron states of the valence and conduction bands, together with
their corresponding wave functions. Using these states and wave
functions, we then calculate the oscillator strengths of the optical
transitions as well as the optical absorption coefficients for the
$\sigma_+$ polarization \cite{Madelung}.

We use the following Luttinger parameters: \mbox{$\gamma_1$=4.14},
\mbox{$\gamma_2$=1.09}, \mbox{$\gamma_3$=1.62}, and
\mbox{$\gamma_{el}$=10.42} ($m_e=0.096\,m_0$) in the QW and in the
barriers. A total discontinuity of bands \mbox{$V_T$=500~meV} and a
valence band offset \mbox{$vbo$=0.4} is assumed, meaning that the
discontinuity of the valence band is \mbox{$V_h$=$V_T\times
vbo$=200~meV} and the discontinuity of the conduction band is
\mbox{$V_{el}$=$V_T\,(1-vbo)$=300~meV} (see Fig.~\ref{figQW}). We
also choose a quantum well of $L_{QW}$=20~\AA, for which the
splitting between the heavy hole $HH_1$ and light hole $LH_1$ energy
states is around 50~meV. Additionally, there is only one heavy hole
state and one light hole state for these parameters.

We assume that the iron island is in a single domain state, with
magnetization $\vec{M}=M_0\vec{e}_x$ pointing in the
\mbox{x-direction} (Fig.~\ref{figQW}). We then calculate the
non-homogeneous field $\vec{B}$ outside the micromagnet by solving
the magnetostatic equations analytically. For instance, when the
distance between the island and the QW is $d$=10~nm, the maximum
value of the magnetic field is $|\vec{B}|_{max}$=1~T. On account of
the large value of $D_x$, the local magnetic field
$\vec{B}(\vec{r})$ can be thought of as a sum of two fields produced
by magnetic charges localized on both poles. On each pole, the
magnetic field is localized in the \mbox{x-direction}. In the
\mbox{y-direction} it extends over the whole width $D_y$. We also
stress that the gradient of the magnetic field \cite{Pulizzi} is as
large as \mbox{2~mT/\AA} for $d$=10~nm, so that a precise
determination of $|\vec{B}|_{max}$ or magnetic field profile is not
trivial even in such simple case of a single domain phase.
In Fig.~\ref{FigRectAbsVsd}, we present the absorption spectrum at
three distances $d$ between the QW and the micromagnet: $d$=10~nm,
$d$=30~nm, and $d$=60~nm. We have fixed the thickness of the
micromagnet and the \mbox{g-factors} of the carriers. The numbers
$nm$ above each peak in Fig.~\ref{FigRectAbsVsd} indicate
transitions between the $n$th electron state and a $m$th hole state.
Each transition line was broadened by a Gaussian function with a
1~meV width. In Fig.~\ref{figQW}, as an example, the transition (11)
is also shown as a dashed vertical line. At $d$=10~nm, the energy
difference between the (11) peak and the MAE, which we call here
binding energy, is around~66~meV. It should be noted that
non-diagonal transitions, e.g. (24), are relatively strong in these
hybrids, in contrast to a normal QW where only diagonal transitions
such (11), (22), etc are strong. With increasing $d$ the binding
energy decreases as the local magnetic field - our "confining agent"
- becomes smaller as the separation between the micromagnet and the
QW increases.

In Fig.~\ref{FigRectAbsVsG} we plot the absorption coefficient for
different $g_e$-factors. The binding energy is a linearly increasing
function of the \mbox{$g_e$-factor} for this range of $g_e$ values.
The effective \mbox{g-factors} are functions of temperature
\cite{Gaj1}. The top panel in Fig.~\ref{FigRectAbsVsG} corresponds
to a standard experiment typically performed at a few Kelvins,
whereas the bottom panel corresponds to a mili-Kelvin temperature
measurement.
As the thickness $D_z$ of the micromagnet increases (now $d$ is
constant), so does the binding energy, as can be seen in
Fig.~\ref{FigRectAbsVsDz}, where we present absorption coefficients
for three thicknesses $D_z$. We can also relate this effect to the
fact that the local magnetic field increases when we increase the
thickness $D_z$ of the island. This behavior persists as long as
$D_z$ is smaller than the in-plane dimensions. When $D_z \in
(150,\,300)$\AA, the binding energy is a linear function of $D_z$.
As stated before, we use a line broadening of~1~meV in all the
calculations. In order to show that our predictions are robust, we
compare in Fig.~\ref{FigRectAbsVsDz} the~1~meV width absorption
curve with another curve where the broadening line width is~4~meV.
Even with such a large line width, the (11) peak is clearly visible.
This is  very promising and indicates that our predictions can be
readily tested under typical experimental conditions.

In conclusion, we calculated the optical response of a hybrid
structure composed of a rectangular micromagnet deposited on top of
a diluted magnetic semiconductor quantum well structure. Our
qualitative and quantitative analysis not only suggests the
possibility localizing spin polarized states in this way, but also
allows us to propose different routes for achieving this
experimentally.  In order to measure spin polarized objects inside
the QW it is necessary to produce as large a local magnetic field as
possible. This can be achieved, for example, by utilizing materials
with high saturation magnetization. Our micromagnet simulations,
together with the energy spectra analysis, show that it is better to
deposit thicker ferromagnetic layers on top of
 the QW, as thicker micromagnets lead to an increased local magnetic field.
As we expected, the growth of high quality QW relatively close to
the ferromagnetic/semiconductor interface is of great importance, as
this enhances localization. Finally, since high values of
\mbox{g-factors} are fundamental for the localization and the
manufacturing of efficient traps, optical localization is expected
to be observed at sub-Kelvin temperatures. Finally, since quasi
one-dimensional states emerge only below the poles of the
micromagnet, spatially resolved techniques such as \mbox{$\mu$-PLE}
or NSOM are required to measure the effects presented in this paper.

This research was supported by the National Science Foundation under
NSF-NIRT Grant No. DMR 02-10519, by the U.S. Department of Energy,
Basic Energy Sciences, under Contract No. W-7405-ENG-36 and by the
Alfred P. Sloand Foundation (B. J.).


\end{document}